\documentclass[showpacs,11pt]{revtex4}

\usepackage{epsfig}
\usepackage{amssymb}

\newcommand{\figurewidth}{6in}

\begin{document}

\title{Preon stars: a new class of cosmic compact objects}

\author{J. Hansson\footnote{e-mail: \tt c.johan.hansson@ltu.se} \& F. Sandin\footnote{e-mail: \tt fredrik.sandin@ltu.se}}

\affiliation{Department of Physics, Lule{\aa} University of Technology, SE-971 87 Lule\aa, Sweden}

\begin{abstract}
In the context of the standard model of particle physics, there is a definite upper limit
to the density of stable compact stars. However, if a more fundamental level of elementary
particles exists, in the form of preons, stability may be re-established beyond this limiting
density. We show that a degenerate gas of interacting fermionic preons does allow for stable
compact stars, with densities far beyond that in neutron stars and quark stars. In keeping with
tradition, we call these objects ``preon stars", even though they are small and light compared
to white dwarfs and neutron stars. We briefly note the potential importance of preon stars in
astrophysics, {\it e.g.}, as a candidate for cold dark matter and sources of ultra-high energy
cosmic rays, and a means for observing them.
\end{abstract}


\maketitle

\section{Introduction}

The three different types of compact objects traditionally considered in astrophysics
are white dwarfs, neutron stars (including quark and hybrid stars), and black holes.
The first two classes are supported by Fermi pressure from their constituent particles.
For white dwarfs, electrons provide the pressure counterbalancing gravity. In neutron
stars, the neutrons play this role. For black holes, the degeneracy pressure is overcome
by gravity and the object collapses indefinitely, or at least to the Planck density.

The distinct classes of degenerate compact stars originate directly from the properties
of gravity, as was made clear by a theorem of Wheeler and collaborators in the mid
1960s~\cite{Harrison}. The theorem states that for the solutions to the stellar structure
equations, whether Newtonian or relativistic, there is a change in stability of one radial
mode of normal vibration whenever the mass reaches a local maximum or minimum as a function
of central density. The theorem assures that distinct classes of stars, such as white
dwarfs and neutron stars, are separated in central density by a region in which there
are no stable configurations.

In the standard model of particle physics (SM), the theory of the strong interaction
between quarks and gluons predicts that with increasing energy and density, the coupling
between quarks asymptotically fades away~\cite{Gross,Politzer}. As a consequence of this
``asymptotic freedom", matter is expected to behave as a gas of free fermions at
sufficiently high densities. This puts a definite upper limit to the density of stable
compact stars, since the solutions to the stellar equations end up in a never-ending
sequence of unstable configurations, with increasing central density. Thus, in the
light of the standard model, the densest stars likely to exist are neutron stars, quark
stars or the potentially more dense hybrid stars~\cite{Gerlach,Glendenning2,Schertler}.
However, if there is a deeper layer of constituents, below that of quarks and
leptons, asymptotic freedom will break down at sufficiently high densities, as
the quark matter phase dissolves into the preon sub-constituent phase.

There is a general consensus among the particle physics community, that something new
should appear at an energy-scale of around one TeV. The possibilities are, {\it e.g.},
supersymmetric particles, new dimensions and compositeness. In this letter we consider
``preon models"~\cite{dSouza,PreonTrinity}, \textit{i.e.}, models in which quarks and
leptons, and sometimes some of the gauge bosons, are composite particles built out of more
elementary preons. If fermionic preons exist, it seems reasonable that a new type of
astrophysical compact object, a {\it preon star}, could exist. The density in preon stars
should far exceed that inside neutron stars, since the density of preon matter must be
much higher than the density of nuclear and deconfined quark matter. The sequence of
compact objects, in order of increasing compactness, would thus be: white dwarfs,
neutron stars, preon stars and black holes.


\section{Mass-radius relations}

Assuming that a compact star is composed of non-interacting fermions with mass $m_f$, the
non-general relativistic (Chandrasekhar) expression for the maximum mass is~\cite{Chandra,Landau}:
\begin{equation}
M \simeq \frac{1}{m_f^2}\left(\frac{\hbar c}{G}\right)^{3/2}.
\label{chandra}
\end{equation}
This expression gives a correct order of magnitude estimate for the mass of a white dwarf and a
neutron star. For quark stars, this estimate cannot be used literally, since the mass of quarks
cannot be defined in a similar way as for electrons and neutrons. However, making the simplifying
assumption that quarks are massless and subject to a `bag constant', a
maximum mass relation can be derived~\cite{Shibaji}. The bag constant is a phenomenological
parameter. It represents the strong interactions that, in addition to the quark momenta,
contribute mass-energy to deconfined quark matter, {\it i.e.}, in the same way as the bag constant
for ordinary hadrons~\cite{bag}. The result in~\cite{Shibaji} is somewhat similar to the
Chandrasekhar expression, but the role of the fermion mass is replaced by the bag constant $B$:
\begin{eqnarray}
M &=& \frac{16 \pi B R^3}{3 c^2}, \label{mchandra} \\
R &=& \frac{3 c^2}{16 \sqrt{\pi G B}}. \label{rchandra}
\end{eqnarray}

For preon stars, one can naively insert a preon mass of $m_f \simeq 1$~TeV/c$^2$ in eq.~(\ref{chandra})
to obtain a preon star mass of approximately one Earth mass (M$_\oplus \simeq 6\times10^{24}$~kg). However,
the energy scale of one TeV should rather be interpreted as a length scale, since it originates from the
fact that in particle physics experiments, no substructure has been found down to a scale of a few hundred
GeV ($\hbar c/$GeV$\,\simeq10^{-18}$~m). Since preons must be able to give light particles, {\it e.g.},
neutrinos and electrons, the ``bare" preon mass presumably is fairly small and a large fraction of the
mass-energy should be due to interactions. This is the case for deconfined quark matter, where the bag
constant contributes more than $10$\% of the energy density. Guided by this observation, and lacking
a quantitative theory for preon interactions, we assume that the mass-energy contribution from preon
interactions can be accounted for by a bag constant. We estimate the order of magnitude for the preon
bag constant by fitting it to the minimum density of a composite electron, with mass $m_e=511$~keV/c$^2$
and ``radius" {$R_{e} \lesssim \hbar c/$TeV$\,\simeq 10^{-19}$~m}. The bag-energy is roughly
$4 B\langle V\rangle$~\cite{bag}, where $\langle V\rangle$ is the time-averaged volume of the bag
(electron), so the bag constant is:
\begin{equation}
	B \simeq \frac{E}{4\langle V\rangle} \gtrsim \frac{3\times 511\,{\rm keV}}{16\pi(10^{-19}\,{\rm m})^3}
	\simeq 10^{4}\,{\rm TeV/fm}^{3} 
	\Longrightarrow\, B^{1/4} \gtrsim 10\,{\rm GeV}.
\end{equation}
Inserting this value of $B$ in eqs.~{(\ref{mchandra}),~(\ref{rchandra})}, we obtain an estimate
for the maximum mass, $M_{max} \simeq 10^2$~M$_\oplus$, and radius, $R_{max} \simeq 1$~m, of a
preon star.

\begin{figure}
\epsfig{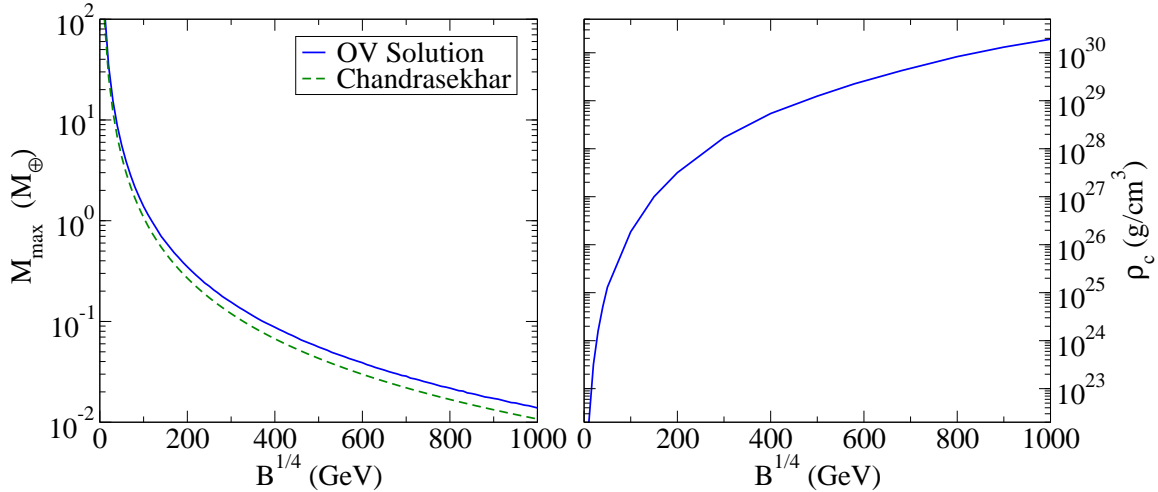}
\caption{The maximum mass and corresponding central density $\rho_c$ of a preon star vs. the bag
constant~$B$. The solid lines represent the general relativistic OV solutions, while the dashed
line represents the Newtonian (Chandrasekhar) estimate. Despite the high central density, the mass
of these objects is below the Schwarzschild limit, as is always the case for static solutions to
the stellar equations. $M_\oplus\simeq6\times10^{24}$~kg is the Earth mass.}
\label{maxmass}
\end{figure}

Since $B^{1/4}\simeq 10$~GeV only is an order of magnitude estimate for the minimum value of $B$,
in the following, we consider the bag constant as a free parameter of the model, with a lower limit
of $B^{1/4}=10$~GeV and an upper limit chosen as $B^{1/4}=1$~TeV.
The latter value corresponds to an electron ``radius" of $\hbar c/10^3\,$TeV$\,\simeq 10^{-22}$~m.
In figs.~\ref{maxmass}~and~\ref{maxr} the (Chandrasekhar) maximum mass and radius of a preon
star are plotted as a function of the bag constant.

Due to the extreme density of preon stars, a general relativistic treatment is necessary.
This is especially important for the analysis of stability when a preon star is subject to
small radial vibrations. In this introductory article we will neglect the effects of rotation
on the composition. Thus, we can use the Oppenheimer-Volkoff (OV) equations~\cite{OV}
for hydrostatic, spherically symmetric equilibrium:
\begin{eqnarray}
\frac{dp}{dr} &=& -\frac{G\left(p+\rho c^2\right)\left(m c^2 + 4 \pi r^3 p\right)}
	{r \left(r c^4 - 2 G m c^2 \right)},  \\
\frac{dm}{dr} &=& 4 \pi r^2 \rho.\label{oveq2}
\end{eqnarray}
Here $p$ is the pressure, $\rho$ the total density and $m=m(r)$ the mass within the radial
coordinate $r$. The total mass of a preon star is $M = m(R)$,
where $R$ is the coordinate radius of the star. Combined with an equation of state (EoS), $p = p(\rho)$,
obtained from some microscopic theory, the OV solutions give the possible equilibrium states
of preon stars.

Since no theory for the interaction between preons yet exists, we make a simple assumption for the
EoS. The EoS for a gas of massless fermions is $\rho c^2 = 3 p$ (see, {\it e.g.},~\cite{Glendenning}),
independently of the degeneracy factor of the fermions. By adding a bag constant $B$, one obtains
$\rho c^2 = 3 p + 4 B$. This is the EoS that we have used when solving the OV equations. The obtained
(OV) maximum mass and radius configurations are also plotted in figs.~\ref{maxmass}~and~\ref{maxr}.


\section{Stability analysis}

A necessary, but not sufficient, condition for stability of a compact star is that
the total mass is an increasing function of the central density ${dM/d\rho_c>0}$~\cite{Glendenning}.
This condition implies that a slight compression or expansion of a star will result
in a less favourable state, with higher total energy. Obviously, this is a necessary
condition for a stable equilibrium configuration. Equally important, a star must be
stable when subject to radial oscillations. Otherwise, any small perturbation would
bring about a collapse of the star.

\begin{figure}
\epsfig{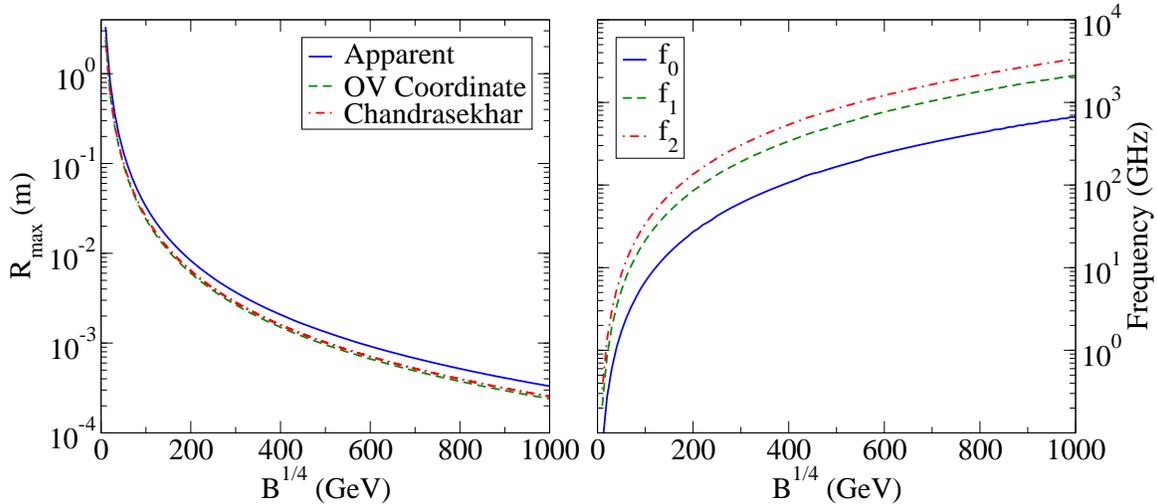}
\caption{The maximum radius and the corresponding first three eigenmode oscillation frequencies
($f_0$,~$f_1$,~$f_2$) vs. the bag constant. The solid line in the left-hand picture is the ``apparent"
radius, $R^\infty = R / \sqrt{1-2 G M / Rc^2}$, as seen by a distant observer. The dashed line
represents the general relativistic coordinate radius obtained from the OV solution, while the
dotted line represents the Newtonian (Chandrasekhar) estimate. Since the fundamental mode $f_0$
is real ($\omega_0^2 > 0$), preon stars with mass below the maximum are stable, for each value of B.}
\label{maxr}
\end{figure}

The equations for the analysis of such radial modes of oscillation are due to
Chandrasekhar~\cite{Chandra2}. An overview of the theory, and some applications, can
be found in~\cite{Misner}. For clarity, we reproduce some of the important points.
Starting with the metric of a spherically symmetric equilibrium stellar model,
\begin{equation}
ds^2 = -e^{2 \nu(r)}dt^2 + e^{2 \lambda(r)}dr^2 + r^2\left(d\theta^2 + \sin^2(\theta)d\phi^2\right),
\end{equation}
and the energy-momentum tensor of a perfect fluid,
$T_{\mu \nu}=(\rho + p)u_\mu u_\nu + p g_{\mu\nu}$,
the equation governing radial adiabatic oscillations can be derived from Einstein's equation.
By making an ansatz for the time dependence of the radial displacement of fluid elements
of the form:
\begin{equation}
\delta r(r,t) = r^{-2} e^{\nu(r)} \zeta(r) e^{i \omega t},
\label{displacement}
\end{equation}
the equation simplifies to a Sturm-Liouville eigenvalue equation for the eigenmodes~\cite{Chandra2,Misner}:
\begin{equation}
\frac{d}{dr}\left( P \frac{d\zeta}{dr}\right) + (Q + \omega^2 W) \zeta = 0.
\label{sturmeq}
\end{equation}
The coefficients $P$, $Q$ and $W$ are~\cite{Misner} (in geometric units where $G=c=1$):
\begin{eqnarray}
P &=& \Gamma r^{-2} p\,e^{\lambda(r) + 3\nu(r)}, \\
Q &=& e^{\lambda(r) + 3\nu(r)} \left[ (p+\rho)^{-1} r^{-2} \left(\frac{dp}{dr}\right)^2
	 - 4 r^{-3}\frac{dp}{dr} - 8\pi r^{-2} p(p+\rho) e^{2 \lambda(r)} \right], \\
W &=& (p+\rho) r^{-2} e^{ 3\lambda(r) + \nu(r)},
\end{eqnarray}
where the adiabatic index $\Gamma$ is:
\begin{equation}
\Gamma = \frac{p+\rho}{p}\left(\frac{\partial p}{\partial \rho}\right)_S.
\end{equation}
The boundary conditions for $\zeta(r)$ are that $\zeta(r)/r^3$ is finite or zero as $r\rightarrow 0$,
and that the Lagrangian variation of the pressure,
\begin{equation}
\Delta p = -\frac{\Gamma p\,e^{\nu}}{r^{2}} \frac{d\zeta}{dr},
\end{equation}
vanishes at the surface of the star.

A catalogue of various numerical methods for the solution of eq.~(\ref{sturmeq}) can
be found in~\cite{Bardeen}. In principle, we first solve the OV equations, thereby
obtaining the metric functions $\lambda(r)$ and $\nu(r)$, as well as $p(r)$, $\rho(r)$
and $m(r)$. Then, the metric functions $\lambda(r)$ and $\nu(r)$ must be corrected for,
so that they match the Schwarzschild metric at the surface of the star (see, {\it e.g.},~\cite{Glendenning}).
Once these quantities are known, eq.~(\ref{sturmeq}) can be solved for $\zeta(r)$ and
$\omega^2$ by a method commonly known as the ``shooting" method. One starts with an
initial guess on $\omega^2$, and integrates eq.~(\ref{sturmeq}) from $r=0$ to the
surface of the star. At this point $\zeta(r)$ is known, and $\Delta p$ can be calculated.
The number of nodes of $\zeta(r)$ is a non-decreasing function of $\omega^2$ (due to
Sturm's oscillation theorem). Thus, one can continue making educated guesses for
$\omega^2$, until the correct boundary condition ($\Delta p = 0$) and number of nodes
are obtained. This method is simple to use when only a few eigenmodes are needed.

\begin{figure}
\epsfig{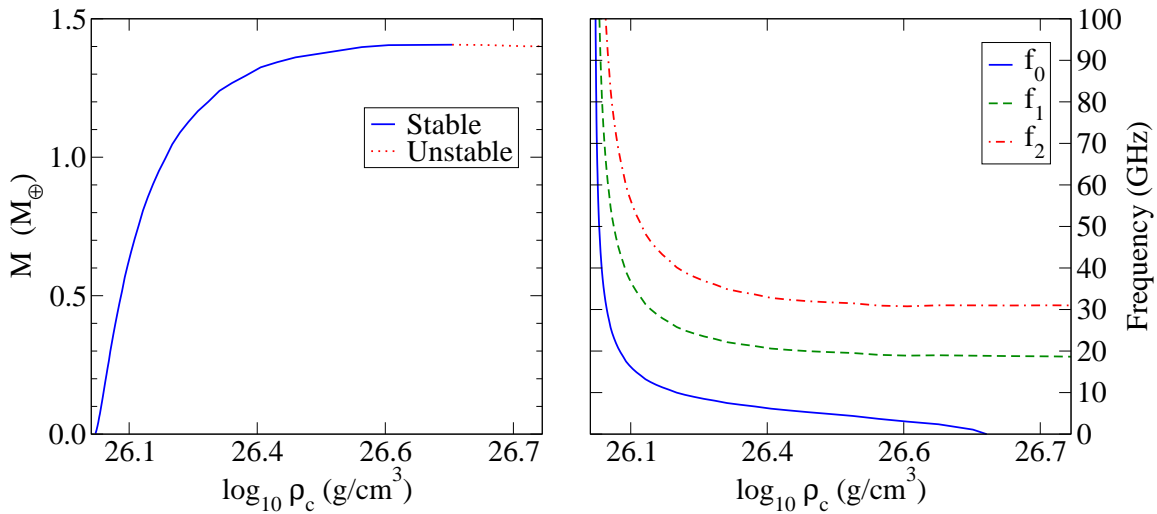}
\caption{The mass and the first three eigenmode oscillation frequencies ($f_0$,~$f_1$,~$f_2$) vs. the
central density $\rho_c$ of preon stars. Here, a fixed value of $B^{1/4}=100$~GeV has been used. For the
maximum mass configuration, the fundamental mode $f_0$ has zero frequency, indicating the onset of
instability. Preon stars with mass and density below the maximum mass configuration of this sequence
are stable.}
\label{b100gev}
\end{figure}

Due to the time dependence in eq.~(\ref{displacement}), a necessary (and sufficient) condition
for stability is that all $\omega_i^2$ are positive. Since $\omega_i^2$ are eigenvalues of a
Sturm-Liouville equation, and governed by Sturm's oscillation theorem, it is sufficient to prove
that the fundamental mode, $\omega^2_0$, is greater than zero for a star to be stable.
In fig.~\ref{b100gev} the first three oscillation frequencies, $f_i=\omega_i/2\pi$, for various
stellar configurations with $B^{1/4}=100$~GeV are plotted.
In agreement with the turning point theorem of Wheeler {\it et al.}~\cite{Harrison},
the onset of instability is the point of maximum mass, as $\omega_0^2$ becomes negative for
higher central densities. Thus, for this value of the bag constant, preon stars are stable
up to the maximum mass configuration.
In order to see if the same is true for other values of $B$, we plot the first three
oscillation frequencies as a function of $B$, choosing the maximum radius configuration
for each $B$. The result can be found in fig.~\ref{maxr}. Indeed, the previous
result is confirmed; all configurations up to the maximum mass are stable.

The eigenmode frequencies for radial oscillations of preon stars are about six orders
of magnitude higher than for neutron stars. This result can also be obtained by making
a simple estimate for the frequency of the fundamental mode. The radius of a preon star
is a factor of $\sim 10^5$ smaller than neutron stars. Hence, if the speed of sound is
similar in preon stars and neutron stars, the frequency would increase by a factor of
$\sim 10^5$, giving GHz frequencies. If the speed of sound is higher in preon stars,
say approaching the speed of light, the maximum frequency is
$\sim 10^8$~m~s$^{-1}/\,0.1$~m~$\simeq 1$~GHz.
Thus, in either case, GHz oscillation frequencies are expected for preon stars.


\section{Potential astrophysical consequences and detection}

If preon stars do exist, and are as small as $10^{-1}-10^{-4}$~m, it is plausible that primordial
preon stars (or ``nuggets") formed from density fluctuations in the early universe. As this material
did not take part in the ensuing nucleosynthesis, the abundance of preon nuggets is not constrained
by the hot big bang model bounds on baryonic matter. Also, preon nuggets are immune to Hawking
radiation~\cite{Hawking} that rapidly evaporates small primordial black holes, making it possible
for preon nuggets to survive to our epoch. They can therefore serve as the mysterious dark matter
needed in many dynamical contexts in astrophysics and cosmology~\cite{DM1,DM2}.

Preon stars born out of the collapse of massive ordinary stars~\cite{Hypernova} cannot contribute
much to cosmological dark matter, as that material originally is baryonic and thus constrained
by big bang nucleosynthesis. However, they could contribute to the dark matter in galaxies. Roughly
$4$\% of the total mass of the universe is in baryonic form \cite{MAP}, but only $0.5$\% is observed
as visible baryons~\cite{PartProp}. Assuming, for simplicity, that all dark matter
$\rho_{DM} = 10^{-25}$~g/cm$^3$ in spiral galaxies, {\it e.g.}, our own Milky Way, is in the form
of preon stars with mass $10^{24}$~kg, the number density of preon stars is of the order of
$10^4$ per cubic parsec ($1$~parsec$\,\simeq 3.1\times 10^{16}$~m). This translates into one preon
star per $10^6$ solar system volumes. However, even though it is not ruled out a priori, the possibility
to form a very small and light preon star in the collapse of a large massive star remains to be more
carefully investigated. In any case, preon nuggets formed in the primordial density fluctuations could
account for the dark matter in galaxies. The existence of such objects can in principle be tested by
gravitational microlensing experiments.

Today there is no known mechanism for the acceleration of cosmic rays with energies above
$\sim 10^{17}$~eV. These so-called ultra-high energy cosmic rays~\cite{UHE} (UHE CR) are
rare, but have been observed with energies approaching $10^{21}$~eV. The sources of UHE CR must,
cosmologically speaking, be nearby ($\lesssim 50$ Mpc $\simeq 150$ million light years) due to the
GZK-cutoff energy $\sim 10^{19}$~eV~\cite{GZK1,GZK2}, since the cosmic microwave background
is no longer transparent to cosmic rays at such high energies. This requirement is very puzzling,
as there are no known sources capable of accelerating UHE CR within this distance.
Preon stars open up a new possibility. It is known that neutron stars, in the form of pulsars,
can be a dominant source of galactic cosmic rays~\cite{Gold}, but cannot explain UHE CR. If for
preon stars we assume, as in models of neutron stars, that the magnetic flux of the parent star is
(more or less) frozen-in during collapse, induced electric fields more than sufficient for the
acceleration of UHE CR become possible. As an example, assume that the collapse of a massive star
is slightly too powerful for the core to stabilize as a $10$~ms pulsar with radius $10$~km, mass
$1.4 M_\odot$ ($M_\odot\simeq 2\times10^{30}$~kg is the mass of our sun) and magnetic field
$10^8$~T, and instead collapses to a preon star state with radius $1$~m and mass $10^2\,M_\oplus$.
An upper limit estimate of the induced electric field of the remaining ``preon star pulsar"
yields $\sim 10^{34}$~V/m, which is more than enough for the acceleration of UHE CR. Also, such strong
electric fields are beyond the limit where the quantum electrodynamic vacuum is expected to break
down, $|\textbf{E}| > 10^{18}$~V/m, and spontaneously start pair-producing particles~\cite{QED}.
This could provide an intrinsic source of charged particles that are accelerated by the electric
field, giving UHE CR. With cosmic ray detectors, like the new Pierre Auger Observatory~\cite{Auger},
this could provide means for locating and observing preon stars.


\section{Conclusions}

In this letter we argue that if there is a deeper layer of fermionic constituents, so-called preons,
below that of quarks and leptons, a new class of stable compact stars could exist. Since no detailed
theory yet exists for the interaction between preons, we assume that the mass-energy contribution
from preon interactions can be accounted for by a `bag constant'. By fitting the bag constant to the
energy density of a composite electron, the maximum mass for preon stars can be estimated to
$\sim 10^{2}$~M$_\oplus$ ($M_\oplus\simeq6\times10^{24}$~kg being the Earth mass), and their maximum
radius to $\sim 1$~m. The central density is at least of the order of $10^{23}$~g/cm$^3$. Preon stars
could have formed by primordial density fluctuations in the early universe, and in the collapse of
massive stars. We have briefly noted their potential importance for dark matter and ultra-high energy
cosmic rays, connections that also could be used to observe them. This might provide alternative means
for constraining and testing different preon models, in addition to direct tests~\cite{PreonTrinity}
performed at particle accelerators.


\section{Acknowledgements}

F. Sandin acknowledges support from the Swedish National Graduate School of Space Technology.
We thank S. Fredriksson for several useful discussions and for reading the manuscript.


\end{document}